\documentclass[prd]{revtex4}

\usepackage{epsfig}
\newcommand{\be}{\begin{equation}}
\newcommand{\ee}{\end{equation}}
\newcommand{\bea}{\begin{eqnarray}}
\newcommand{\eea}{\end{eqnarray}}

\newcommand{\nn}{\nonumber \\}
\newcommand{\e}{{\rm e}}

\begin{document}

\tolerance=5000

\title{Dark Energy: Vacuum Fluctuations, the Effective Phantom Phase,
\\   and Holography}
\author{E. Elizalde$^{1}$\footnote{e-mail: elizalde@ieec.fcr.es}, S.
Nojiri$^{2}$\footnote{e-mail: nojiri@nda.ac.jp,
snojiri@yukawa.kyoto-u.ac.jp}, S. D. Odintsov$^{3}$\footnote{e-mail:
odintsov@ieec.uab.es also at TSPU, Tomsk}, and Peng
Wang$^{4}$\footnote{e-mail:
pewang@eyou.com}}
\affiliation{$^{1}$Consejo Superior de Investigaciones Cient\'{\i}ficas
(ICE/CSIC) \\  Institut d'Estudis Espacials de Catalunya (IEEC) \\
Campus UAB, Fac. Ciencies, Torre C5-Par-2a pl \\ E-08193 Bellaterra
(Barcelona) Spain \\
$^2$Department of Applied Physics, National Defence Academy,
Hashirimizu Yokosuka 239-8686, Japan \\
$^3$Instituci\`o Catalana de Recerca i Estudis Avan\c{c}ats (ICREA) and
Institut d'Estudis Espacials de Catalunya (IEEC/ICE),
Edifici Nexus, Gran Capit\`a 2-4, 08034 Barcelona, Spain \\
$^4$Physics Dept., Nankai Univ., Tianjin,  300071, Peoples
Republic of China}


\begin{abstract}
We aim at the construction of dark energy models without exotic matter
but with a phantom-like equation of state (an effective phantom phase).
The first model we consider is decaying vacuum cosmology
where the fluctuations of
the vacuum are taken into account. In this case, the phantom cosmology
(with an effective, observational $\omega$ being less than $-1$ )
    emerges even for the case of a real
dark energy with a physical equation of state parameter $\omega$ larger
than $-1$. The second proposal is a generalized holographic model,
    which is produced by the presence of an infrared cutoff. It
also leads to an effective phantom phase, which is not a transient
one as in the first model. However, we show that quantum effects are
able to prevent its evolution towards a Big Rip singularity.

\end{abstract}


\maketitle

\section{Introduction}



Recent observational constraints obtained for the dark energy
equation of state (EOS) indicate that our Universe is probably in a
superaccelerating phase (dubbed as ``phantom cosmology"), i.e. $\dot
H>0$ (see \cite{ywang} and references therein). When dark energy is
modeled as a perfect fluid, it can be easily found that
in order to drive a
phantom cosmology, the dark energy  EOS should satisfy
$\omega_d\equiv\rho_d/p_d<-1$ (phantom)
      \cite{caldwell}. The classical phantom has already many similarities
with a quantum field \cite{NO}. However, it violates important
energy conditions and all attempts to consider it as a (quantum)
field theory show that it suffers from instabilities both from a
quantum mechanical and from a gravitational viewpoint
\cite{carroll-phantom}. Thermodynamics of this theory are also quite
unusual \cite{brevik,odintsov-de}: the corresponding entropy must be
negative or either negative temperatures must be introduced in order
to obtain a positive entropy. The typical final state of a phantom
universe is a Big Rip singularity \cite{bigrip} which occurs within
a finite time interval. (Note that  quantum effects coming from
matter may slow up or even prevent the Big Rip singularity
\cite{odintsov,odintsov-de}, which points towards a transient
character of the phantom era).

In the light of such problems, we feel that it would be very
interesting to construct a new scenario where an (effective) phantom
cosmology could be described through more consistent dark energy
models with $\omega_d\ge -1$. Recently, some proposals along this
line have already appeared. An incomplete list of them includes the
consideration of modified gravity \cite{abdalla} or generalized
gravity \cite{carroll-trick}, taking into account quantum effects
\cite{Onemli}, non-linear gravity-matter couplings \cite{couple},
the interaction between the photon and the axion \cite{csaki},
two-scalar dark energy models \cite{odintsov, hu}, the braneworld
approach (for a review, see \cite{sahni}, however, FRW brane
cosmology may not be predictive \cite{boundary}), and
several others.

In this paper we continue the construction of more consistent
dark energy models, with an effective phantom phase but without
introducing the phantom field, what is realized by using a number
of considerations from holography and vacuum
fluctuations. In the next section, a dark energy model which
includes vacuum fluctuations effects (a decaying vacuum cosmology)
is considered. With reasonable assumptions about the vacuum decaying
law, the FRW equations are modified. Then, it is shown that the
effective dark energy
EOS parameter, which is constrained by astrophysical observations,
is less than $-1$. However, even if the dark energy EOS parameter
is actuall bigger than $-1$, the modified FRW expansion, or in
other words, a decaying vacuum can trick us into thinking about
some exotic matter with $w$ less than $-1$.
Hence, such dark energy model does not need the exotic phantom
matter, its effective phantom phase is transient and it does not
contain a Big Rip singularity.
In section 3 another holographic dark energy model which uses the
existence of an infrared cutoff, identified with a particle or future
horizon, is constructed. It generalizes some previous attempts in this
direction and leads to modified FRW dynamics. Such model also
contains an effective phantom phase (without introduction of a
phantom field), which leads to Big Rip type singularity.
However, taking into account matter quantum effects may again prevent
(or moderate) the Big Rip emergence, in the same way as in
Refs.~\cite{odintsov, odintsov-de}. A summary and outlook is given in
the Discussion section.

\section{Dark energy and vacuum fluctuations}

When considering the energy of the vacuum, there are two
conceptually different contributions: the energy of the vacuum state
and the energy corresponding to the \emph{fluctuations} of the
vacuum. More precisely, the first is the eigenvalue of the
Hamiltonian acting on the vacuum state: $E=\langle0|H|0\rangle$,
while the second is the dispersion of the Hamiltonian:
$E_\Delta=\langle0|H^2|0\rangle-(\langle0|H|0\rangle)^2$.

In constructing the dark energy as a scalar field
(``quintessence") \cite{quint}, only the energy of the vacuum state
is actually considered, and the vacuum fluctuations are neglected.
Note that, while the potential energy of the scalar field
corresponds to the energy of the vacuum state, the kinetic energy
density does not correspond to the energy of the vacuum
fluctuations. The kinetic energy corresponds to the energy of the
excited states. Actually, ignoring vacuum fluctuations may not be a
well justified assumption. It was ignored for quintessence models,
just for simplicity.

Thus, confronted with the difficulties of building a phantom
universe from quintessence alone, it is natural to consider the
sector which was ignored before. In this paper, we will show that
the ignored sector can really help to resolve the problems above.
The key point is that, when considering those two contributions to
the vacuum energy in a cosmological setup, their energy density
changes with the cosmological expansion in different ways.

First, if  dark energy is modeled as a scalar field $\phi$,
interacting only with itself and with gravity \cite{quint}, then it
behaves just like an ordinary perfect fluid satisfying the
continuity equation
\begin{equation}
\dot\rho_d+3H(\rho_d+p_d)=0 , \label{1}
\end{equation}
where
\begin{equation}
\rho_d={1\over2}\dot\phi^2+V(\phi) ,\label{1b}
\end{equation}
and
\begin{equation}
p_d={1\over2}\dot\phi^2-V(\phi) .\label{1c}
\end{equation}
Thus
\begin{equation}
\omega_d=\frac{{1\over2}\dot\phi^2-V(\phi)}{{1\over2}\dot\phi^2+V(\phi)}\ge-1.\label{1d}
\end{equation}
When comparing dark energy models with observations, it is often
assumed that $\omega_d=\text{const.}$ (this is reasonable since
allowing EOS to vary will increase the number of parameters to
constrain). Then from equation (\ref{1}) it follows that
\begin{equation}
\rho_d \propto (1+z)^{3(1+\omega_d)} .\label{2}
\end{equation}

On the other hand, to find the energy corresponding to the
fluctuations of the vacuum, we must have a regularization scheme in
order to handle the divergent expression \cite{brustein}. Since the
regularization scheme must be Lorentz invariant (as is the case,
e.g. of dimensional regularization in curved spacetime), it turns
out that the final result must be of the form
$T_{\Lambda\mu\nu}=\rho_\Lambda g_{\mu\nu}$ even if $\rho_\Lambda$
is time-varying due to the time variation of the IR cutoff
\cite{pad}. Note that \emph{mathematically}, $\rho_\Lambda
g_{\mu\nu}$ is exactly the energy-momentum of a perfect fluid with
$p=-\rho$. Then from the energy-momentum conservation law, we find
that the vacuum fluctuations will interact with matter
\cite{brostein, wang, horvat}
\begin{equation}
\dot\rho_m+3H\rho_m=-\dot\rho_\Delta ,\label{3}
\end{equation}
where $\rho_m$ is the energy density of nonrelativistic matter and
$\rho_\Delta$ is the energy density of vacuum fluctuations.
Following \cite{wang}, we will call a cosmological model which includes
the effects of vacuum fluctuations a ``decaying vacuum cosmology"
since, as can be seen from (\ref{3}), during the expansion of the
Universe, owing to the fact that $\rho_\Delta$ is decreasing, new
matter is created from the vacuum.

It is clear that Eq.~(\ref{3}) and the Friedmann equation are not
enough to describe a decaying vacuum cosmology. We should also
specify a vacuum decaying law. There are lots of papers on various
proposals concerning how $\rho_\Delta$ will change with cosmological
time (see \cite{wang} and references therein). However, recently it
has been proposed  \cite{wang} that from (\ref{3}) and a simple
assumption about the form of the modified matter expansion rate, the
vacuum decaying law can be constrained to a two-parameter family.
Thus, we can think about the observational consequences of a
decaying vacuum cosmology, even if we do not understand the physics
underlying the decaying law. More precisely, the main (and rather
natural) assumption is that the matter expansion rate be of the form
\begin{equation}
\rho_m=\rho_{m0}a^{-3+\epsilon}\ . \label{4}
\end{equation}
where $\rho_{m0}$ is the present value of $\rho_m$. Actually, this
assumption is valid in all the existing models of a decaying vacuum
cosmology. An immediate simple observation is that one must have
$\epsilon \le1$. Otherwise, the Universe would expand with an
acceleration in the matter dominated era, which is excluded by the
observations of SNe Ia, that our Universe expanded with deceleration
before redshift $z\sim 0.5$ \cite{ywang}. Actually, it is expected
that $\epsilon\ll 1$ since, so far, there has been no observational
evidence about an anomalous dark matter expansion rate.

      From (\ref{3}) and (\ref{4}), we can find that $\rho_\Delta$ is
given by
\begin{equation}
\rho_{\Delta}=\tilde\rho_{\Delta0}+{\epsilon\rho_{m0}\over
3-\epsilon}a^{-3+\epsilon}\ ,\label{5}
\end{equation}
where $\tilde\rho_{\Delta0}$ is an integration constant. In
\cite{wang} it was  shown that all the existing vacuum decaying laws
can be described by a suitable choice of $\epsilon$ and
$\tilde\rho_{\Delta}$. In our current framework, the term
$\tilde\rho_{\Delta}$ can be absorbed into the dark energy density,
so in the following discussion, we will simply assume
$\tilde\rho_{\Delta}=0$. The FRW equation describing a Universe
consisting of cold dark matter, dark energy and vacuum fluctuations
is
\begin{equation}
H^2={1\over 3M_P^2}(\rho_m+\rho_d+\rho_\Delta) .\label{5b}
\end{equation}

It is important to note that due to the modified dark matter
expansion rate, the physical dark energy EOS $\omega_d$ is no longer
the one directly seen in supernova observations. To compare our
model with observations, we can adopt the framework of effective
dark energy EOS \cite{linder} that can unify quintessence, modified
gravity, and decaying vacuum cosmology into one and a single
framework. The effective dark energy EOS is defined as \cite{linder}
\begin{equation}
\omega_{eff}=-1+{1\over 3}\frac{d\ln\delta H^2(z)}{d\ln(1+z)}\
,\label{6}
\end{equation}
where $\delta H^2\equiv H^2-\rho_{m0}(1+z)^3/(3M_P^2)$ characterizes
any contribution to the cosmic expansion in addition to the standard
cold dark matter. It is $\omega_{eff}$ that is the real quantity
which is constrained in cosmological observations. The explicit form
of $\delta H^2$ in the above model is
\begin{equation}
{\delta H^2(z)\over H_0^2}={3\Omega_{m0}\over
3-\epsilon}(1+z)^{3-\epsilon}-\Omega_{m0}(1+z)^3+\Omega_{d0}(1+z)^{3(1+\omega_d)},
\label{7}
\end{equation}
from which we can find the effective dark energy EOS:
\begin{equation}
\omega_{eff}(z)=\omega_d+\frac{\omega_d(1+z)^3-{3\omega_d+\epsilon\over
3-\epsilon}(1+z)^{3-\epsilon}}{{3\over3-\epsilon}(1+z)^{3-\epsilon}-(1+z)^3
+{\Omega_{d0}\over\Omega_{m0}} (1+z)^{3(1+\omega_d)}} . \label{eEOS}
\end{equation}
     From this one can see that, ignoring the decay of the vacuum, i.e. with
$\epsilon=0$, then  $\omega_{eff}=\omega_d$.

\begin{figure}
       \includegraphics[width=0.8\columnwidth]{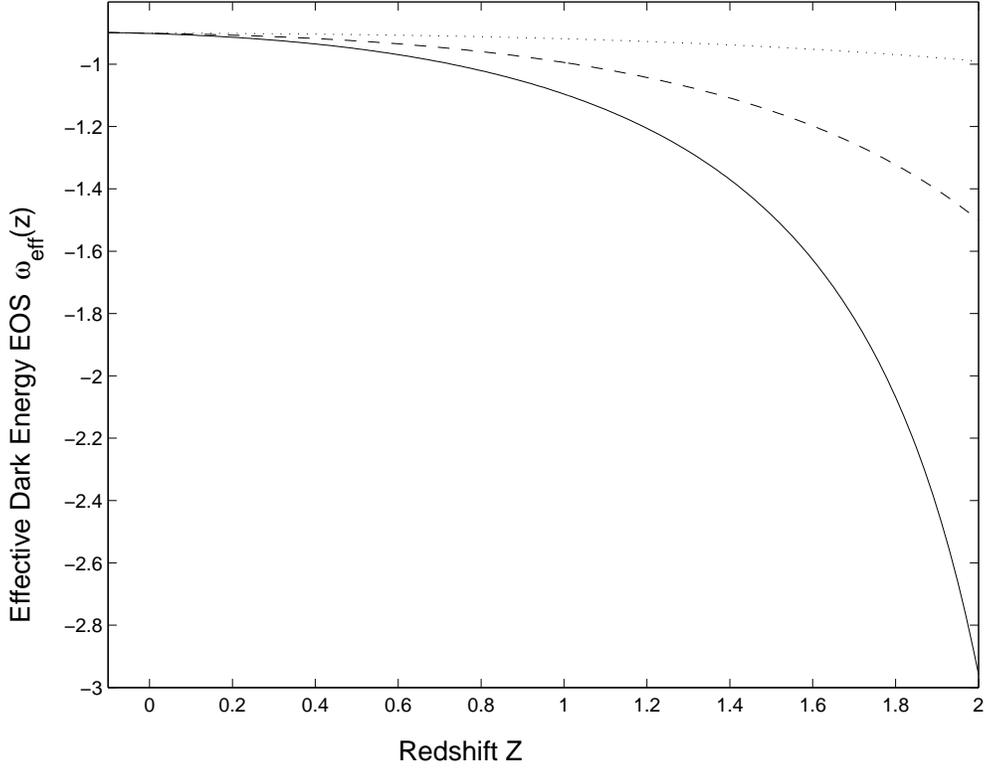}
       \caption{Evolution of the effective dark energy EOS as given by equation
(\ref{eEOS}) with
       $\omega_d=-0.9$. Dotted, dashed and solid lines
      correspond to $\epsilon=0.01, 0.05, 0.1$, respectively.}\label{F1}
\end{figure}

FIG.~\ref{F1} shows the evolution of $\omega_{eff}$ up to redshift
2 for $\epsilon=0.01, \omega_d=0.9$ from top to bottom. It is easy
to see that $\omega_{eff}=\omega_d$ for small $z$ and negative $z$.
As mentioned above, in current works  constraining $\omega_{eff}$,
the most reliable  assumption is that $\omega_{eff}$ is a constant
\cite{ywang}. Thus, what is actually constrained in those works is
the average value of the effective EOS
$\bar\omega_{eff}\equiv\int\Omega_{eff}(a)\omega_{eff}(a)da/\int\Omega_{eff}(a)da$
\cite{limin}, which from FIG.~\ref{F1} is smaller than $-1$ if
averaged up to redshift 2 (the current upper bound on observable
supernova) for roughly $\epsilon>0.05$. Thus, we have constructed a
model which can drive a phantom cosmology with a physical EOS larger
than $-1$. In this picture, $\bar\omega_{eff}<-1$ is not the result
of exotic behavior of the dark energy, but the modified expansion
rate of the CDM, i.e. a decaying vacuum can trick us into thinking
that $\omega_{eff}<-1$. Under the assumption of a constant EOS, the
current constraint on $\bar\omega_{eff}$ from SNe Ia data is
$-4.36<\bar\omega_{eff}<0.8$ at $95\%$ C.L. \cite{ywang}. Then
$\epsilon<0.1$ is obviously consistent with current constraints on
dark energy EOS. Moreover, it is easy to see from FIG.~\ref{F1} that
$\bar\omega_{eff}$ will be more and more negative when the redshift
gets larger. Thus, if more high redshift SNe Ia data become
available in the future, the interaction between vacuum and
nonrelativistic matter predicts that we will get a more negative
value for $\bar\omega_{eff}$.

It is also interesting to note that in the current model, the
arguments leading to a Big Rip of the Universe in a phantom
cosmology \cite{bigrip} no longer apply. While the current expansion
rate is superaccelerating, the Universe will be driven by only the
healthy $\omega_X\ge-1$ dark energy in the future. The Universe is
now superaccelerating just because non-relativistic matter is of the
same order of magnitude as dark energy, and thus the effects of the
nonstandard expansion rate of nonrelativistic matter cannot be
ignored. This conclusion is supported by recent analysis on the
construction of a sensible quantum gravity theory in cosmological
spacetime \cite{bousso}. Actually, it is easy to see that the
trouble with de Sitter space discussed in \cite{bousso} will be
avoided in a superaccelerating cosmology: the total amount of
Hawking radiation due to the cosmological horizon is infinite and
thus it will thermalize the observer. Furthermore, since the horizon
is decreasing in a superaccelerating cosmology, black holes will
dominate the spectrum of thermal fluctuations in late time
\cite{bousso}. An observer may be swallowed by a horizon sized black
hole before he/she can reach the Big Rip singularity. Thus, it is
doubtful that the appearance of Big Rip singularity makes sense in
the quantum gravity theory. This is also supported by a direct
account of the quantum gravity effects moderating or preventing the
Big Rip \cite{odintsov, odintsov-de}.

In the above discussion, the simplest model of dark energy
characterized by a constant EOS is considered with the assumption
that the vacuum decays only to dark matter. Let us now discuss a
more general case. Let the vacuum fluctuation energy decay into both
dark matter and dark energy. Specifically, we will consider a dark
energy with EOS proposed in \cite{odintsov-de} (see, also
\cite{stefa, odintsov}),
\begin{equation}
p_d=-\rho_d-A\rho_d^\alpha ,\label{de}
\end{equation}
where $A$ and $\alpha$ are constant. This can be viewed as
characterizing the minimum deviation of the dark energy EOS from
that of a cosmological constant.
\begin{figure}
       \includegraphics[width=0.8\columnwidth]{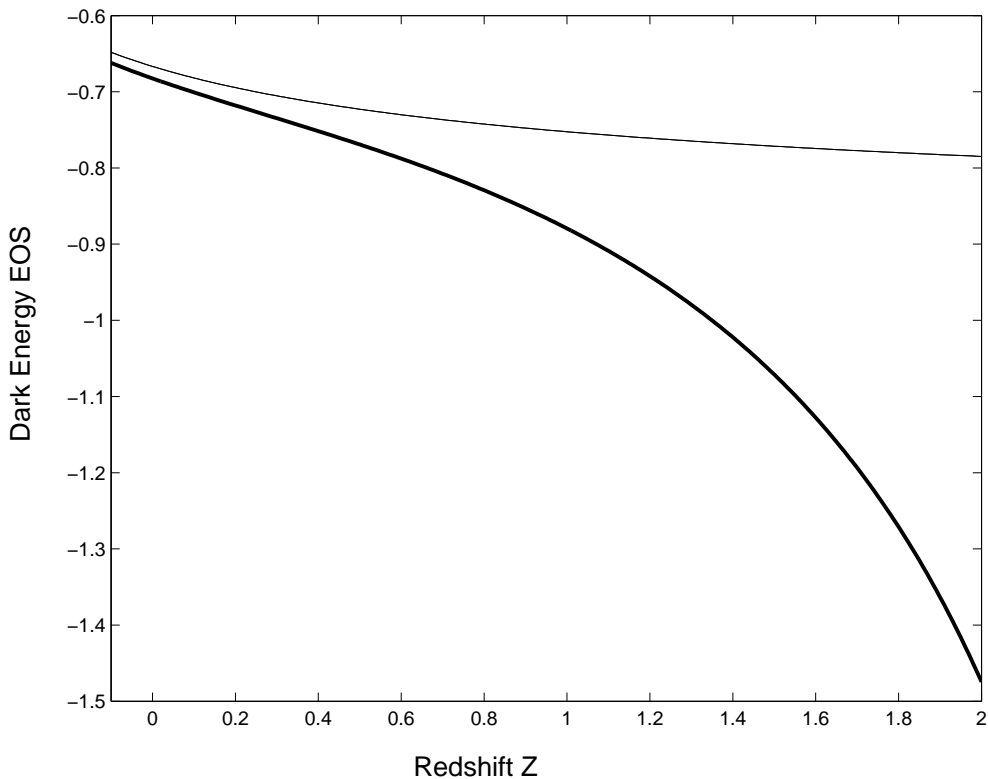}
       \caption{Evolution of the physical (thin line) and effective (thick line)
dark energy EOS as given by equations (\ref{pheos}) and
(\ref{eEOS2}).}\label{F1b}
\end{figure}
When considering the effect of vacuum fluctuations, from the total
energy-momentum conservation, one gets
\begin{equation}
\dot\rho_{d}+\dot\rho_m+3H(\rho_{d}+p_{d}+\rho_m)=-\dot\rho_\Delta
.\label{9}
\end{equation}

Now one cannot proceed further without specifying a vacuum decaying
law. Currently,  the following law can be proposed
\begin{equation}
\rho_\Delta=\epsilon M_P^2H^2 .\label{law}
\end{equation}

This law can be deduced from the results of \cite{brustein}, in
which  the vacuum fluctuations of a spatially finite quantum system,
e.g. free scalar field, have been evaluated. The most interesting
conclusion in this computation is that the energy corresponding to
the vacuum fluctuations is proportional to the area of the spatial
boundary of the system. In a cosmological setting, it is reasonable
to view the Hubble horizon as the spatial boundary of the quantum
theory describing matter in the Universe.

     From equation (\ref{law}) and the Friedmann equation, it can be
shown that Eq.~(\ref{9}) can be written as
\begin{equation}
\dot\rho_{d}+\dot\rho_m+(3-\epsilon)H(\rho_{d}+p_{d}+\rho_m)=0,
\label{10}
\end{equation}
from which we find that
\begin{equation}
\rho_{d}=\rho_{d0}[1-(3-\epsilon)\tilde A(1-\alpha)\ln
(1+z)]^{1/(1-\alpha)} ,\ \ \rho_m=\rho_{m0}(1+z)^{3-\epsilon}
,\label{11}
\end{equation}
where $\tilde A\equiv A\rho_{d0}^{\alpha-1}$. For $\epsilon=0$, i.e.
in the absence of vacuum decaying effects, (\ref{11}) reduces to the
expression found in \cite{stefa, odintsov}.

Now, one can find the effective dark energy EOS by exactly the same
procedure as above:
\begin{equation}
\omega_{eff}=-1+
\frac{(1+z)^{3-\epsilon}-(1+z)^3-{(3-\epsilon)\tilde
A\Omega_{d0}\over 3\Omega_{m0}}(1-(3-\epsilon)\tilde
A(1-\alpha)\ln
(1+z))^{\alpha/(1-\alpha)}}{{3\over3-\epsilon}(1+z)^{3-\epsilon}-(1+z)^3+{\Omega_{d0}\over\Omega_{m0}}
(1-(3-\epsilon)\tilde A(1-\alpha)\ln (1+z))^{1/(1-\alpha)}},
\label{eEOS2}
\end{equation}
When $\epsilon=0$, this reduces to the physical equation of state,
\begin{equation}
\omega_d={p_d\over\rho_d}=-1-\tilde A(1-3\tilde A(1-\alpha)\ln
(1+z))^{-1}.\label{pheos}
\end{equation}

FIG.~\ref{F1b} shows the evolution of $\omega_{eff}$ and $\omega_d$
up to redshift 2 for $\tilde A=-1/3, \alpha=1/2, \epsilon=0.1$.
Hence, it is obvious that while $\omega$ for the physical dark
energy EOS is larger than $-1$, with account of vacuum fluctuations,
     for the dark energy EOS is smaller than $-1$. Thus, we have constructed
a vacuum fluctuation dark energy model where the phantom phase is
transient and there is {\it no need} to introduce exotic matter (a
phantom)  to comply with the observational data.

\section{Holographic dark energy and the phantom era}

The fluctuations of the vacuum energy are caused by quantum effects.
It is expected, that the quantum corrections to the theories coupled
with gravity should be constrained by the holography. The holography
requires an infrared cutoff $L_\Lambda$, which has dimensions of
length and could be dual to the ultraviolet cutoff. Hence, the
fluctuations of the vacuum energy are naturally proportional to the
inverse square of the infrared cutoff. This points to the connection
of dark energy based on fluctuations of the vacuum energy with the
holographic dark energy. Our purpose here will be to construct the
generalization of the holographic dark energy model which naturally
contains the effective phantom phase.

Let us start from  the holographic dark energy model which is a
generalization of the model in \cite{Li} (see also
\cite{Myung1,Myung2,FEH}). Denote the infrared cutoff by
$L_\Lambda$, which has a dimension of length. If the holographic
dark energy $\rho_\Lambda$ is given by, \be \label{H1}
\rho_\Lambda=\frac{3c^2}{\kappa^2 L_\Lambda^2}\ , \ee with a
numerical constant $c$, the first FRW equation
$(H^2/\kappa^2)=\rho_\Lambda$ can be written as \be \label{H2}
H=\frac{c}{L_\Lambda}\ . \ee Here it is assumed that $c$ is positive
to assure the expansion of the universe. The particle horizon $L_p$
and future horizon $L_f$ are defined by \be \label{H3} L_p\equiv a
\int_0^t\frac{dt}{a}\ ,\quad L_f\equiv a \int_t^\infty \frac{dt}{a}\
. \ee For the FRW metric with the flat spacial part: \be \label{H4}
ds^2 = -dt^2 + a(t)^2\sum_{i=1,2,3}\left(dx^i\right)^2\ . \ee
Identifying $L_\Lambda$ with $L_p$ or $L_f$, one obtains the
following equation: \be \label{H5}
\frac{d}{dt}\left(\frac{1}{aH}\right)=\pm \frac{c}{a}\ . \ee Here,
the plus (resp. minus) sign corresponds to the particle (resp.
future) horizon. Assuming \be \label{H6} a=a_0 t^{h_0}\ , \ee it
follows that \be \label{H7} h_0=1\mp c\ . \ee Then, in the case
$L_\Lambda=L_f$, the universe is accelerating ($h_0>0$ or $w=-1 +
{2}/{3h_0}<-1/3$). When $c>1$ in the case $L_\Lambda=L_p$, $h_0$
becomes negative and the universe is shrinking. If the theory is
invariant under the change of the direction of time, one may change
$t$ with $-t$. Furthermore by properly shifting the origin of time,
we obtain, instead of (\ref{H6}), \be \label{H8} a=a_0\left(t_s -
t\right)^{h_0}\ . \ee This tells us that there will be a Big Rip
singularity at $t=t_s$. Since  the direction of time is changed, the
past horizon becomes a future like one: \be \label{H8b} L_p\to
\tilde L_f \equiv a \int_t^{t_s}\frac{dt}{a}=a \int_0^\infty
\frac{da}{Ha^2}\ . \ee

Note that if  $L_\Lambda$ is chosen as a future horizon in
(\ref{H3}), the deSitter space \be \label{dS1}
a=a_0\e^{\frac{t}{l}}\quad \left(H=\frac{1}{l}\right) \ee can be a
solution. Since $L_f$ is now given by $L_f=l$, we find that the
holographic dark energy  (\ref{H1}) is given by
$\rho_\Lambda=\frac{c^2}{\kappa^2 l^2}$. Then when $c=1$, the first
FRW equation $\frac{3}{\kappa^2} H^2 = \rho_\Lambda$ is identically
satisfied. If $c\neq 1$, the deSitter solution is not a solution. If
we choose $L_\Lambda$ to be the past horizon, the deSitter solution
does not exist, either, since the past horizon $L_p$  (\ref{H3}) is
not a constant: $L_p=(1 - \e^{\frac{t}{l}} )/l$.

In \cite{Hsu}, it has been argued that, when we choose $L_\Lambda$ to be
the particle horizon $L_p$, the obtained energy density  gives a natural 
value
consistent with the observations but the parameter $w$ of the equation of 
the state vanishes since $L_p$ should behave as $a(t)^{3/2}$. 
That seems to contradict with the observed value.
In the present paper, we wish to generalize the holographic  model to have, hopefully,
more freedom and to be more successful in cosmology.
As we will see in this section, there may be many possiblilities for the 
choice of the infrared cutoff $L_\Lambda$ which are more consistent with observations
(giving correct cosmology without above problems).

In general, $L_\Lambda$ could be a combination (a function) of both,
$L_p$, $L_f$. Furthermore, if the span of life of the universe is
finite, the span $t_s$ can be an infrared cutoff. If the span of
life of the universe is finite, the definition of the future horizon
$L_f$  (\ref{H3}) is not well-posed, since $t$ cannot go to
infinity. Then, we may redefine the future horizon as in (\ref{H8b})
\be \label{H8c} L_f\to \tilde L_f \equiv a
\int_t^{t_s}\frac{dt}{a}=a \int_0^\infty \frac{da}{Ha^2}\ . \ee
Since there can be many choices for the infrared cutoff, one may
assume $L_\Lambda$ is the function of $L_p$, $\tilde L_f$, and
$t_s$, as long as these quantities are finite: \be \label{H9}
L_\Lambda=L_\Lambda\left(L_p, \tilde L_f, t_s\right)\ . \ee As an
example, we consider the model \be \label{H10} \frac{L_\Lambda}{c} =
\frac{2t_s \left(\frac{L_p + \tilde L_f}{\pi t_s}\right)^2}{\left\{
1 + \left(\frac{L_p + \tilde L_f}{\pi t_s}\right)^2\right\}^2}\ ,
\ee which leads to the solution: \be \label{H11}
H=\frac{1}{2}\left(\frac{1}{t} + \frac{1}{t_s - t}\right)\ ,\quad
\mbox{or}\quad a=a_0 \sqrt{\frac{t}{t_s - t}}\ . \ee In fact, one
finds \be \label{H12} L_p+\tilde L_f = a \int_0^\infty
\frac{da}{Ha^2} = a \int_0^{t_s} \frac{dt}{a}=\pi t_s
\sqrt{\frac{t}{t_s - t}}\ , \ee and therefore \be \label{H13}
\frac{c}{L_\Lambda}=\frac{1}{2}\left(\frac{1}{t} + \frac{1}{t_s -
t}\right)=H\ , \ee which satisfies (\ref{H2}).

We should note that any kind of finite large quantity could be
chosen as the infrared cutoff. For instance, $L_\Lambda$ could
depend on $dL_p/dt$ and/or $d{\tilde L}_f/dt$. The following model
may be considered: \be \label{H14} \frac{\tilde
c}{L_\Lambda}=\frac{1}{L_p + \tilde L_f}\left(\frac{dL_p}{dt} +
\frac{d\tilde L_f}{dt}\right)\ , \ee with the assumption that \be
\label{H15} \tilde c\equiv \int_0^\infty
\frac{da}{Ha^2}=\int_0^{t_s}\frac{dt}{a} \ee is finite. Since
$\tilde c$ is a constant and $L_p + \tilde L_f=a\tilde c$, from
(\ref{H14}), we find that (\ref{H2}) is trivially satisfied.
Therefore, in the model (\ref{H14}), as long as $\tilde c$ in
(\ref{H15}) is finite, an arbitrary FRW metric is a solution of
(\ref{H2}).

The next step is to consider the combination of the holographic dark
energy and other matter  with the following EOS \be \label{H16}
p=-\rho - f(\rho)\ . \ee The holographic dark energy is given by
(\ref{H1}) with $L_\Lambda=L_p$, $L_f$ in (\ref{H3}), or $\tilde
L_f$ in (\ref{H8b}) or (\ref{H8c}). Then from the first FRW equation
\be \label{H16b} \pm \frac{d}{dt}\left\{\frac{c}{a}\left(H^2 -
\frac{\kappa^2}{3}\rho\right)^{-\frac{1}{2}}\right\} =\frac{1}{a}\ .
\ee First, we consider the case when $f(\rho)=-(1+w)\rho$ with
constant $w$, that is, \be \label{H17} p=w\rho\ . \ee Then, by using
the conservation of the energy \be \label{H18} 0=\frac{d\rho}{dt} +
3H\left(\rho + p\right)\ , \ee it follows \be \label{H19} \rho =
\rho_0 a^{-3(1+w)}\ . \ee By combining (\ref{H16b}) with
(\ref{H19}), one finds a solution \bea \label{H20} && a=\left\{
\begin{array}{lll}  a_0 t^{\frac{2}{3(1+w)}} \quad & \mbox{when} \quad & w>-1
\\
a_0 \left(t_s -t\right)^{\frac{2}{3(1+w)}} \quad & \mbox{when} \quad & w<-1 \\
\end{array} \right. \nn
&& \rho_0 a_0^{3(1+w)} =
\frac{3}{\kappa^2}\left\{\frac{2(1-c)}{3(1+w)} + c\right\}
\left\{\frac{2(1+c)}{3(1+w)} - c\right\} \ . \eea $c$ is assumed to
be non-negative. Hence, in order for $\rho_0 a_0^{3(1+w)}$
(\ref{H20}) to be positive,  the constraint for the solution
(\ref{H20}) appears: \be \label{H21} -\frac{1}{3}-
\frac{2}{3c}<w<-\frac{1}{3} + \frac{2}{3c}\ . \ee Then, if $w<-1$,
it follows $c<1$.

For another EOS (see also the previous section) \be \label{H22}
f(\rho)=f_0\rho^\alpha\ , \ee
      the conservation of the energy (\ref{H18}) gives
\be \label{H23} \rho = \left\{ 3f_0 (1-\alpha)\ln
\frac{a}{a_0}\right\}^{\frac{1}{1-\alpha}}\ . \ee Defining a new
variable $q$ by \be \label{H24} q\equiv \ln \frac{a}{a_0}\ , \ee
since \be \label{H25} \frac{d}{dt}=aH\frac{d}{da} = H\frac{d}{dq}\ ,
\ee Eq.(\ref{H16b}) can be rewritten as \be \label{H26} \pm c
\frac{d}{dq}\left\{\e^{-q}\left(H^2 -
\frac{\kappa^2}{3}\rho\right)^{-\frac{1}{2}}\right\}
=\frac{\e^{-q}}{H}\ . \ee Thus, Eq.(\ref{H23}) gives \be \label{H27}
\rho = \rho_1 q^\beta\ ,\quad \rho_1\equiv \left\{ 3f_0
(1-\alpha)\right\}^{\frac{1}{1-\alpha}}\ , \quad \beta\equiv
\frac{1}{1-\alpha}\ . \ee As we are interested in the case when $a$,
and therefore $q$, is large, we assume that $H$ behaves as $H\sim
h_1 q^\beta$, when $q$ is large. From (\ref{H26}), one gets \be
\label{H29} \left(1-c^2\right)h_1^2 = \frac{\kappa^2}{3}\rho_1\ ,
\ee which requires $c^2\leq 1$. At the same time, \be \label{H30}
a\propto \e^{\left(\frac{h_1}{1 - \frac{\beta}{2}}
t\right)^{\frac{1}{1 - \frac{\beta}{2}}}} \quad \mbox{or} \quad
\e^{\left(-\frac{h_1}{1 - \frac{\beta}{2}} \left(t_s -
t\right)\right)^{\frac{1}{1 - \frac{\beta}{2}}}}\ . \ee Hence, if $1
- \frac{\beta}{2}<0$ or $1/2<\alpha<1$, $a$ diverges at $t=t_s$,
which is similar to a Big Rip singularity\cite{bigrip}.

Let us now consider the contribution of quantum corrections. Quantum
corrections are important in the early universe when $t\sim 0$ and
curvatures are large. However, the quantum corrections are important
near the Big Rip singularity, where the curvatures are large around
the singularity at $t=t_s$. Since quantum corrections usually
contain the powers of the curvature or higher derivative terms, such
correction terms play important role near the singularity. It is
interesting to take into account the back reaction of the quantum
effects near the singularity\cite{odintsov, odintsov-de}.
Note that holographic dark energy could be caused by
non-perturbative quantum gravity. However, it is not  easy to
determine its explicit form. As it has been shown above, the Big Rip
may occur also in a holographic dark energy model. On the other
hand, the contribution of the conformal anomaly as a back reaction
near the singularity could be determined in a rigorous way. In the
situation when the contribution from the matter fields dominates and
one can neglect the quantum gravity contribution, quantum effects
could be responsible for the change of the FRW dynamics
non-perturbatively. Hence, as in other dark energy models
\cite{odintsov, odintsov-de} the Big Rip singularity could be indeed
moderated or even prevented.

The conformal anomaly $T_A$ has the following form: \be \label{OVII}
T_A=b\left(F+\frac{2}{3}\nabla^2 R\right) + b' G + b''\nabla^2 R\ ,
\ee where $F$ is the square of a 4d Weyl tensor and $G$ is a
Gauss-Bonnet curvature invariant. In general, with $N$ scalar,
$N_{1/2}$ spinor, $N_1$ vector fields, $N_2$ ($=0$ or $1$) gravitons
and $N_{\rm HD}$ higher derivative conformal scalars, the
coefficients $b$ and $b'$ are given by \bea \label{bs}
\hspace*{-2.2em} && b=\frac{N +6N_{1/2}+12N_1 + 611 N_2 - 8N_{\rm
HD}}{120(4\pi)^2}\,, \nn \hspace*{-2.2em} &&
b'=-\frac{N+11N_{1/2}+62N_1 + 1411 N_2 -28 N_{\rm
HD}}{360(4\pi)^2}\,. \eea
     $b>0$ and $b'<0$ for the usual matter except for higher derivative
conformal scalars.
We should note that $b''$ can be shifted by a finite renormalization of the
local counterterm $R^2$, so  $b''$ can be arbitrary.

In terms of the corresponding energy density $\rho_A$ and the pressure $p_A$,
$T_A$ is given by $T_A=-\rho_A + 3p_A$.
Using the energy conservation law in the FRW universe:
\be
\label{CB1}
\dot{\rho}_A+3 H\left(\rho_A + p_A\right)=0\,,
\ee
one may delete $p_A$ as
\be
\label{CB2}
T_A=-4\rho_A -\dot{\rho}_A/H \,.
\ee
This gives the following expression for $\rho_A$:
\bea
\label{CB3}
\rho_A&=& -\frac{1}{a^4} \int_{t_0}^t d t\, a^4 H T_A \nn
&=&  -\frac{1}{a^4} \int_{t_0}^t d t\, a^4 H \Bigl[-12b \dot{H}^2 + 24b'
(-\dot{H}^2 + H^2 \dot{H} + H^4)  \nn
& &- (4b + 6b'')\left(\dddot{H} + 7 H \ddot{H} + 4\dot{H}^2 + 12 H^2 \dot{H}
\right) \Bigr]\,.
\eea

The next step is to consider the FRW equation \be \label{H31}
\frac{3}{\kappa^2}H^2=\rho_\Lambda + \rho_A\ . \ee The natural
assumption is that the scale factor $a$ behaves as (\ref{H6}) when
$t\sim 0$ or as (\ref{H8}) when $t\sim t_s$. Then $H^2$ and
$\rho_\Lambda$ behave as $t^{-2}$ or $\left(t_s - t\right)^{-2}$ but
$\rho_A$ behaves in a more singular way as $t^{-4}$ or $\left(t_s -
t\right)^{-4}$. Hence, the power law behavior cannot be a solution
of (\ref{H31}). Instead, the spacetime approaches the deSitter space
as in (\ref{dS1}). From (\ref{H31}) one finds \be \label{H32}
\frac{3(1-c^2)}{\kappa^2 l^2}=-\frac{6 b'}{l^4}\ . \ee When $c<1$,
which corresponds to the singularity (\ref{H6}) of the early
universe, Eq.(\ref{H32}) has solutions \be \label{H33}
\frac{1}{l^2}=0\ ,\ \ -\frac{1-c^2}{2 b' \kappa^2}\ . \ee Thus, the
singularity in the early universe may be stopped by quantum effects.
The whole universe could be generated by the quantum effects. On the
other hand, if $c>1$ the only consistent solution of (\ref{H32}) is
\be \label{H34} H=\frac{1}{l}=0\ . \ee This shows that the universe
becomes flat. Therefore due to the quantum matter back-reaction, the
Big Rip singularity could be avoided.
Hence, unlike to the model of previous section,
the effective phantom phase for holographic dark energy leads to Big Rip
which may be moderated (or prevented) by other effects.

\section{Discussion}

Summing up, we have considered in this paper two dark energy models
both of which contain an
effective phantom phase of the universe evolution without the
actual need to introduce a scalar phantom field.
Both models, which can be termed as decaying vacuum cosmology and
(generalized) holographic dark
energy, have a similar origin related with quantum considerations.
However, the details of the two models are quite different.
For instance, the decaying vacuum cosmology is caused by vacuum
fluctuations and, under a very reasonable assumption about the
decaying law, the effective
phantom phase is transient, no Big Rip occurs. Moreover, the observable
(effective) EOS parameter being the phantom-like one to comply with
observational data is different from the real dark energy EOS
parameter, which may be bigger than $-1$. At the same time,
holographic dark
energy is motivated by AdS/CFT-like holographic considerations
(emergence  of an infrared cutoff).
Even without the phantom field, the effective phantom phase there
leads to a Big Rip singularity, which most probably can be avoided by
taking into account quantum and quantum gravity effects.

It looks quite promising that one can add to the list of existing
dark energy models with phantom-like EOS (some of them have
been mentioned in the introduction) two additional models which
  exhibit the interesting
new property that there is no need to introduce exotic matter
explicitly (as this last is known to
violate the basic energy conditions). Definitely,
one may work out these theories in more detail. However, as usually
happens, it is most probable that the truth lies in between
the models at hand, and
that a realistic dark energy ought to be constructed as some
synthesis of the existing approaches which seem to possess reasonable
properties. New astrophysical
data are needed for the resolution of this important
challenge of the XXI century.

\section*{Acknowledgments}

This research has been supported in part by the Ministry of
Education, Science, Sports and Culture of Japan under grant
n.13135208 (S.N.), by RFBR grant 03-01-00105, by LRSS grant
1252.2003.2 (S.D.O.), and  by DGICYT (Spain), project
BFM2003-00620 and SEEU grant PR2004-0126 (E.E.).

\end{document}